\documentclass[twocolumn]{revtex4}
\usepackage{graphicx}
\begin{document}
\title{Entangled states of light}
\author{S.J. van Enk}
\affiliation{Bell Labs, Lucent Technologies\\
600-700 Mountain Ave,
Murray Hill, NJ 07974}
\affiliation{Institute for Quantum Information,\\
California Institute of Technology,\\
Pasadena, CA 91125
}

\date{\today}

\begin{abstract}
These notes are more or less a faithful representation of
my talk at the Workshop on ``Quantum Coding and Quantum Computing''
held at the University of Virginia. As such it is an introduction
for non-physicists to the topics of the quantum theory of light and 
entangled states of light.
In particular, I discuss the photon concept
and what is really entangled in an
entangled state of light (it is not the photons!).
Moreover, I discuss an example that highlights
the peculiar behavior of entanglement in an infinite-dimensional
Hilbert
space.
\end{abstract}

\maketitle
\section{Light}
Thanks to a certain class of T-shirts \cite{T} we all know that light
is described by Maxwell's equations. These equations 
determine how the electric and magnetic fields vary in space and time, given
electric currents and charges.
Fortunately, even if there are no sources, 
that is, no currents and no charges, 
there are still
nontrivial solutions. It is those solutions that describe light waves.
One way to express such a light wave is to write 
down the electric field vector
as a function of position $\vec{r}$ and time $t$,
\begin{equation}
\vec{E}(\vec{r},t)=
\vec{\epsilon}
\exp i\big(\vec{k}\cdot\vec{r}-\omega t\big)+{\rm complex}
\;{\rm conjugate}.
\end{equation}  
This wave corresponds to a monochromatic plane
wave with frequency $\omega$,
propagating in the direction
of $\vec{k}$, and with a polarization $\vec{\epsilon}$.
These quantities are not independent as they satisfy
\begin{equation}
|\vec{k}|c=\omega;\,\,\vec{k}\cdot\vec{\epsilon}=0.
\end{equation}
Here $c$ is the speed of, what else, light.
The general solution to the source-free Maxwell equations is a sum 
(or an integral, depending on whether we allow $\vec{k}$ to take
continuous values) 
of 
plane waves, according to
\begin{eqnarray}\label{sol}
\vec{E}(\vec{r},t)&=&\sum_{\vec{k}}\sum_{\vec{\epsilon}\perp\vec{k}}
A_{\vec{k},\vec{\epsilon}}\,
\vec{\epsilon}
\exp i\big(\vec{k}\cdot\vec{r}-\omega t\big)\nonumber\\
&&+{\rm complex}
\;{\rm conjugate},
\end{eqnarray}
where $A_{\vec{k},\vec{\epsilon}}$ are arbitrary complex
coefficients, called amplitudes, that
determine  the intensity of the light beam.

The Maxwell equations and hence its solutions (\ref{sol})
are ``classical'', as       
they do not describe any quantum effects.
So, what changes if we are to use a quantum theory of light?
Some may think that the frequency $\omega$ becomes quantized, that is, 
takes on only discrete values. Others may think the polarization vector 
$\vec{\epsilon}$
becomes a quantum object. Pessimists may think that the Maxwell equations
are no longer valid and perhaps polarization and frequency are no longer well
defined. But none of these expectations are correct.
What really happens when one quantizes 
the electromagnetic field is the following:
Just as position and momentum of a particle become 
operators acting on some complex Hilbert space 
when one quantizes the particle, so do
the electric (and magnetic) fields become operators in a quantum theory of 
light. It turns out that the Maxwell equations keep in fact the same form. 
So, the solutions (\ref{sol}) are also still valid,
except that ``complex conjugate'' becomes  ``Hermitian conjugate.'' 

The  difference with the classical solutions (\ref{sol}) is that the 
amplitudes $A_{\vec{k},\vec{\epsilon}}$ become operators.
We typically write the amplitude as the product of a number 
(with the dimension of an electric field), and a dimensionless operator 
$a_{\vec{k},\vec{\epsilon}}$. For each value of $(\vec{k},\vec{\epsilon})$,
called a {\em mode},
there is such an operator $a$ 
(leaving out the subscripts for ease of notation). 
That operator acts on a complex Hilbert space. In the 
case of the electromagnetic field, that Hilbert space is infinite-dimensional.
It is spanned by states $|n\rangle$ for $n=0,1,2\ldots$, and the operator $a$
acts as
\begin{equation}\label{a}
a|n\rangle=\sqrt{n}|n-1\rangle.
\end{equation}
The number $n$ is interpreted as the number of photons in the 
corresponding mode. 
Thus the state $|2\rangle$ is a state with two photons in a particular mode.
The operator $a$ annihilates one photon, and is called the annihilation operator.
Similarly, the Hermitian conjugate operator, $a^{\dagger}$ acts as
\begin{equation}\label{ad}
a^{\dagger}|n\rangle=\sqrt{n+1}|n+1\rangle,
\end{equation}
and is called the creation operator.
The fact that the number states $|n\rangle$ start with 
$n=0$ is not just a convenient convention: 
Eq.~(\ref{a}) shows that $a$ acting on the state $|0\rangle$ 
gives nothing, zero, not even a state. 
We cannot have a state with $-1$ photons. 

For later use we note that if we define $N=a^\dagger a$, we 
have in fact managed to define the number operator,
\begin{equation}\label{N}
N|n\rangle=n|n\rangle,
\end{equation}
that counts the number of photons in a mode.
\subsection{Entangled states of light}
To make an entangled state of light we need (at least) two modes. 
Suppose we have two modes, denoted by $A$ and $B$. 
Modes $A$ and $B$ may correspond to modes with
the same frequency and the same wave vector, 
but different (orthogonal) polarizations.
Or they may have the same polarization and the same frequency, 
but propagate in a different (not necessarily orthogonal) direction. 
In any case, the Hilbert space $H_{A,B}$
associated with the
two modes is the tensor product $H_{A,B}=H_A\otimes H_B$ of the Hilbert spaces 
of each of the modes.

Let us look at a simple example of a state in $H_{A,B}$:
\begin{equation}\label{ex1}
|\Psi\rangle_{A,B}=[|0\rangle_A|1\rangle_B+|1\rangle_A|0\rangle_B]/\sqrt{2}.
\end{equation}
Although properly we should have written $|0\rangle_A\otimes |1\rangle_B$ etc., we used here the 
physicists' convention to be lazy.
This state is entangled because it cannot be written as a 
product of two states of the modes
$A$ and $B$. Thus neither mode is in a well-defined pure state of itself.
Loosely speaking, we could say there is one photon, and it is either in mode $A$ or in mode $B$.
This way of speaking misses some essential points though. The state (\ref{ex1})
expresses more than just a classical joint probability distribution to find the photon
in modes $A$ and $B$. In particular, if modes $A$ and $B$ are spatially separated, then 
the statistics of measurement outcomes on the state (\ref{ex1}) can in general
not be obtained from any classical
joint probability distribution without superluminal communication.
Since physicists don't like superluminal communication, this expresses a serious difference
between quantum and classical mechanics. 
[Note that this does not imply that entanglement can be used for 
superluminal communication, it is just that a classical local model 
cannot explain quantum mechanics.]
This basically is the content of Bell's theorem.
I refer to Bell's book \cite{bell} for the details.
\subsection{What is entangled?}  
We found that the state (\ref{ex1}) is entangled. But what is entangled
with what? 
One might think photons are entangled with each other.
That is indeed not the most stupid answer one can give, as after all light
consists of photons. Nevertheless, 
it is not the photons 
that are entangled. In fact, there is only one photon in the state 
(\ref{ex1}). 
Instead, modes $A$ and $B$ are entangled with each other. 
The number of photons just determines the 
states of the modes.

Now, however, this leads to a slight problem. 
After all, nobody tells us how to define our modes. We can take any 
complete set of functions to expand our electric field in, as in Eq.~(\ref{sol}),
and to each member of that set corresponds a mode. 
So unlike in the case of, say, atoms in a trap, we are more or less free to define what we mean by our systems.
With atoms in a trap, the atoms are clearly the systems, but for light modes it is a lot less clear.
For instance, if modes $A$ and $B$ correspond to horizontal and vertical polarization, respectively 
(with all other mode variables, the wave vector and the frequency, the same), then the entangled state (\ref{ex1})
turns out to be equal to a state with one diagonally polarized photon. 
Formally, by defining new annihilation operators
$a'_{\pm}=(a_A\pm a_B)/\sqrt{2}$, we find that there is one photon in mode $'+'$ and no photons in mode $'-'$.
However, that state we would never call entangled. 

On the other hand, suppose modes $A$ and $B$ correspond to spatially separated modes, mode $A$ is here, 
mode $B$ is there. The $\pm$ modes can theoretically still be defined. However, in practice the modes 
``here plus there''
and ``here minus there'' are useless. 
In particular, noone would know how to directly measure a photon in such a mode.
(Of course, one can always indirectly measure such modes by recombining the modes so that 
they end up in the same location. But that's cheating as one explicitly changes the modes and thereby the 
entanglement.) Thus, when the modes $A$ and $B$ are spatially separated no redefinitions of those modes are useful.
So we decree that it makes sense to talk unambiguously about entanglement between two 
modes $A$ and $B$ only if the modes are spatially
separated. If the modes are in the same location we do allow redefinitions of the modes and 
thereby of the entanglement.
\subsection{Some remarks}
The importance attached to nonlocality for the issue of defining entanglement is consistent with
several other results: first of all, 
there is above-mentioned Bell's theorem. Second, 
there has been a lot of interest in relativistic aspects of entanglement of both light and particles, 
such as electrons. One conclusion \cite{scudo}, 
in particular, is that the entanglement between different degrees of 
freedom of one particle (namely, the spin and the momentum)  is not Lorentz-invariant. That is,
the amount of entanglement present within that one particle depends on the velocity of the observer. The reason is that different observers don't agree on what
the spin and momentum degrees are. 
This is similar to having different observers disagreeing
on how the modes of the electromagnetic field are defined.
When talking about two particles in different positions, no such problems arise.
Third, it has been shown \cite{spreeuw} that many quantum properties of light, even including 
entanglement,  
can be simulated by using just classical light beams. However,
this is only so if the entanglement is not nonlocal. 

The point that entanglement is between modes and not between photons was made in several different papers \cite{pz,em},
all coming out at roughly the same time. Apart from making that same point,
the papers do have different ideas otherwise.
In particular, my own attempt \cite{em} at a paper discusses by how much
redefinitions of modes can change the entanglement. For example, in the simple
state  (\ref{ex1}), the smallest and largest possible amounts of entanglement one can get
are just zero, and one unit of entanglement, respectively.
There are other simple examples where the entanglement cannot be 
transformed away completely, and also examples where the maximum 
entanglement possible is probably more than what one would have expected at first and even second glance.
  
Finally, I also note that the importance of what can actually be measured
for the definition of entanglement was discussed in a more general
and more precise context in \cite{barnum}. This was also discussed at the 
Workshop by Lorenza Viola.
\section{Entangled coherent states} 
This Section serves a number of purposes. First, it shows an explicit 
nontrivial (as opposed to (\ref{ex1})) example of an entangled state.
Second, that example will display some curious behavior that is known to 
occur in principle
in infinite-dimensional Hilbert spaces, but for which there was no more or less
practical example known. 
 
The easiest way to generate a light beam with nice quantum properties is to switch on a laser.
The quantum state of the light beam is a so-called coherent state 
[disregarding some subtleties that are of no interest here]. 
For any complex number $\alpha$ it is defined as
\begin{equation}\label{coh}
|\alpha\rangle:=\exp(-|\alpha|^2/2)
\sum_{n=0}^\infty \frac{\alpha^n}{\sqrt{n!}}|n\rangle,
\end{equation}
in terms of the number states $|n\rangle$. The easiest way to produce a nontrivial state of two modes, is to take 
a laser beam and split it on a beam splitter.
Annoyingly, though, one can never get an entangled state out of these two easy operations.
If we split two coherent states on a beam splitter we just get two new coherent states with different amplitudes,
but nothing more. 
In fact, more generally speaking, with linear optics one cannot create entanglement from coherent states.
The reason is that all linear optics elements just lead to linear transformations of operators $a_i$ of modes $i$.
Namely, the general transformation is of the form
\begin{equation}\label{trans}
a'_j=\sum_i U_{ij}a_j,
\end{equation}
with $U_{ij}$ a unitary matrix. Since a coherent state (\ref{coh}) is an eigenstate of $a$, and since
the transformations
(\ref{trans}) can only transform eigenstates of the operators $a_i$ into eigenstates of new mode operators
$a'_j$, coherent states are always transformed into coherent states.

Thus, in order to create entanglement from a coherent state one needs nonlinear optics.
One particular type of nonlinearity I consider here is the Kerr nonlinearity. It is characterized by a 
Hamiltonian of the form
\begin{equation}\label{H}
H=\hbar \chi a^{\dagger2}a^2,
\end{equation}
where $\hbar$ is Planck's constant.
The Hamiltonian determines the evolution in time of any state, with
 the evolution operator as a function of time $t$ given by
\begin{equation}
U(t)=\exp(-it H/\hbar).
\end{equation}
Starting from a coherent state $|\alpha\rangle$ at time $t=0$
the state evolves thus as
\begin{equation}
|\psi(t)\rangle=\exp(-it H/\hbar)|\alpha\rangle.
\end{equation}
The evolution operator can be rewritten in terms of the number operator
defined in (\ref{N}),
\begin{equation}
U(\tau)=\exp(-i \tau a^{\dagger2}a^2)=
\exp(-i \tau N(N-1)),
\end{equation}
where $\tau=\chi t$ is a dimensionless time.
The operator $U$ becomes periodic in $N$ with period $M$ 
(that is, it becomes invariant under $N\rightarrow N+M$) 
at times $\tau=\pi/M$ if $M$ is an odd integer.
This implies one can write down Fourier series as follows
[following Ref.~\cite{tara}]
\begin{equation}\label{odd}
\exp\big( \frac{-i\pi}{M}N(N-1)  \big)=\sum_{q=0}^{M-1} f_q^{(o)}\exp
\big( \frac{-2i\pi q}{M}N\big).
\end{equation}
Similarly, for even values of $M$ one has
\begin{equation}\label{even}
\exp\big( \frac{-i\pi}{M}(N+M)^2  \big)=
\exp\big( \frac{-i\pi}{M}N^2  \big),
\end{equation}
so that we can expand
\begin{equation}\label{Feven}
\exp\big( \frac{-i\pi}{M}N^2  \big)=\sum_{q=0}^{M-1} f_q^{(e)}\exp
\big( \frac{-2i\pi q}{M}N\big).
\end{equation}
The coefficients $f_q$ are not explicitly evaluated in 
Ref.~\cite{tara}, but one can 
actually derive them (see \cite{multi} for more details),
\begin{eqnarray}\label{MM}
f_q^{(o)}&=&\frac{1}{\sqrt{M}}\exp\big(\frac{\pi iq(q+1)}{M}  \big)
\exp\big(\frac{-\pi iK(K+1)}{M}  \big),\nonumber\\
f_q^{(e)}&=&\frac{1}{\sqrt{M}}\exp\big(\frac{\pi iq^2}{M} \big) \exp(-\pi i/4),
\end{eqnarray}
where in the first line $K$ is such that $M=2K+1$ for odd $M$.
The expansions (\ref{odd}) and (\ref{even}) are
 useful as it becomes easy to calculate the effect of the 
evolution opertor on a coherent state. The reason is the simple relation
\begin{equation}
\exp(i\phi N)|\alpha\rangle=|\alpha \exp(i\phi)\rangle.
\end{equation}
Thus, if one starts with a coherent state  
at time $\tau=0$, 
this then immediately leads to the following time evolution under $U$:
\begin{eqnarray}
U(\pi/M)|\alpha\rangle
&=&\sum_{q=0}^{M-1}f_q^{(o)}|\alpha\exp(-2\pi iq/M)\rangle\nonumber\\
{\rm for}\,\,M\,\,{\rm odd},\nonumber\\
U_A(\pi/M)|\alpha\rangle
&=&\sum_{q=0}^{M-1}f_q^{(e)}|\alpha\exp(\pi i(1-2q)/M)\rangle\nonumber\\
{\rm for}
\,\,M\,\,{\rm even}\end{eqnarray}
If one subsequently takes these states and splits them on a 50/50 beamsplitter with the vacuum, the output state is an entangled state of the form
\begin{equation}\label{ent2}
|\Phi_M\rangle=\sum_{q=0}^{M-1}f_q^{(o)} |\beta\exp(-2\pi iq/M)\rangle
 |\beta\exp(-2\pi iq/M)\rangle,
\end{equation}
for $M$ odd 
with $\beta=\alpha/\sqrt{2}$, and something similar for even $M$.

The state (\ref{ent2}) is entangled. A measure for how much 
entanglement there 
is in a state of two modes can be obtained after first
writing the state in a standard form, 
the so-called Schmidt-decomposition form. 
There is a unique 
(up to some trivial transformations) way to write a bipartite state as
\begin{equation}\label{S}
|\Psi\rangle=\sum_{q=0}^{M-1} \sqrt{p}_q |\psi_q\rangle|\phi_q\rangle,
\end{equation}
where the $p_q$ are positive real numbers with $\sum_q p_q=1$
 (so they can be interpreted as probabilities), and with 
the states $|\psi_q\rangle$ for $q=0\ldots M-1$ all orthogonal to each other,
and the same for $|\phi_q\rangle$. 
The entanglement is then given by
\begin{equation}
E=-\sum p_q\log_2p_q,
\end{equation}
which some may recognize as the Shannon entropy of the probability distribution $\{p_q\}$.
If we have $M$ terms in (\ref{S}) then the entanglement can at most be $\log_2 M$.
Now the state (\ref{ent2}) is almost written in the Schmidt form (\ref{S}). It is only ``almost'', as
I had not told you yet 
that the coherent states are not orthogonal, as can be checked directly
from the definition.
However, for large values of $|\alpha|$ the different
coherent states do become orthogonal, 
so in that limit the entanglement in the state (\ref{ent2}) is
$E=\log_2 M$.
\begin{figure}
\includegraphics[scale=0.4]{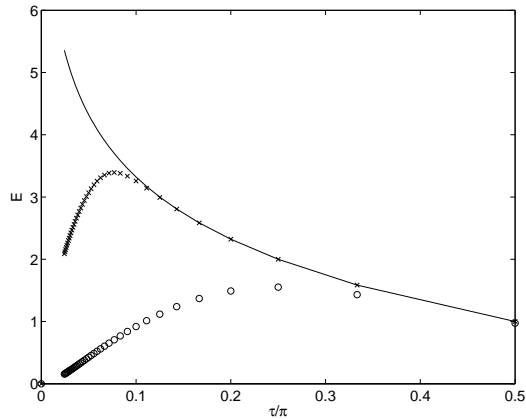}
\caption{Entanglement for the states (\ref{ent2}) as a function of $\tau$, for 
various values of $|\alpha|^2$: $|\alpha|^2=1$ for the bottom curve (circles), $|\alpha|^2=10$
for the middle curve (crosses), and $|\alpha|^2\rightarrow\infty$ for the top curve (solid line).}
\end{figure}

Now here is a peculiar effect: we get more entanglement with increasing $M$, that is, if we remember we 
generated this state after a time $\tau=\pi/M$, with {\em decreasing} time. 
So the shorter the nonlinear Hamiltonian (\ref{H}) acts, 
the more entanglement we get (and we started with no entanglement).
How can this possibly be right?
The answer lies hidden in the assumption of large $\alpha$: 
for increasing $M$ the different coherent states are only orthogonal when 
$|\alpha/M|\gg 1$. So, with increasing $M$ one needs ever larger values of $\alpha$. 
Now the energy in a coherent state
is proportional to $|\alpha|^2$, so the energy needed to make $\log_2 M$ units of entanglement grows as $M^2$.
Now it is known \cite{wehrl}
that entanglement in infinite dimensions is not a continuous function of the state. 
But, if one restricts oneself to states with an upper bound on the total energy, then that function is continuous.
Here too, the peculiar behavior of entanglement disappears 
when we {\em fix} the value of $\alpha$ and then calculate the entanglement as a function of time $\tau$.
For small $\tau$ the entanglement will be small: the above-mentioned argument that 
the entanglement should be large does not hold for fixed $\alpha$ as the coherent states 
become very non-orthogonal for large $M$.
This is illustrated in Fig.~1, where the entanglement as a function of $\tau$ is plotted
for 2 different values of $\alpha$ together with the maximum possible amount of entanglement
(only reached for $|\alpha|\rightarrow\infty$).

A final question to be answered is the following. 
Even though we know now that there is no real contradiction arising
from the fact that in principle a large amount of entanglement can 
be created after a short interaction time, one may wonder whether that effect can be useful for creating lots of entanglement without being bothered too much by decoherence. 
After all, a shorter interaction time implies in general
 less decoherence. 
If you fund my research, please stop reading now.
Unfortunately, there is a catch. 
One only creates those large amounts of entanglement in short times
if the amplitude $\alpha$ is large. That in turn leads to more decoherence. 
The more photons there are in a state, the easier they are lost.
A more precise analysis shows that the increasing decoherence indeed 
seriously counteracts 
the potential increase of entanglement. Thus,
it seems one is not better off using very short interaction times to create entanglement with the Kerr interaction. 
On that happy note, I end.
\section{Acknowledgments}
I thank Olivier Pfister for having been a great host and Karen Klintworth for 
having taken care of all the annoying but important details.

\end{document}